\documentclass[journal=jacsat,manuscript=article]{achemso}

\usepackage[version=3]{mhchem} 

\author{Omar Nuruzade}
\affiliation[Khazar]
{Department of Life Sciences, Khazar University, 41 Mahsati Street, Baku, AZ1096, Azerbaijan}
\alsoaffiliation[French-Azerbaijani University]
{Department of Oil and Gas Engineering, French-Azerbaijani University, 183 Nizami Street, Baku, AZ1000, Azerbaijan}

\author{Elshan Abdullayev}
\affiliation[Khazar]
{Department of Life Sciences, Khazar University, 41 Mahsati Street, Baku, AZ1096, Azerbaijan}

\author{Valentina Erastova}
\email{valentina.erastova@ed.ac.uk}
\affiliation[UoE]
{School of Chemistry, University of Edinburgh, Joseph Black Building, David Brewster Road, King’s Buildings, Edinburgh, EH9 3FJ, UK}

\title[Apigenin and smectite clay - the effect of pH and cations]
  { Organic-mineral interactions under natural conditions -- a computational study of flavone adsorption on smectite clay }

\abbreviations{MD, Api, MMT, SWy }

\keywords{Apigenin, montmorillonite, smectite clay, molecular modelling, flavonoids}

\begin{document}

\begin{tocentry}

\includegraphics[scale=0.4]{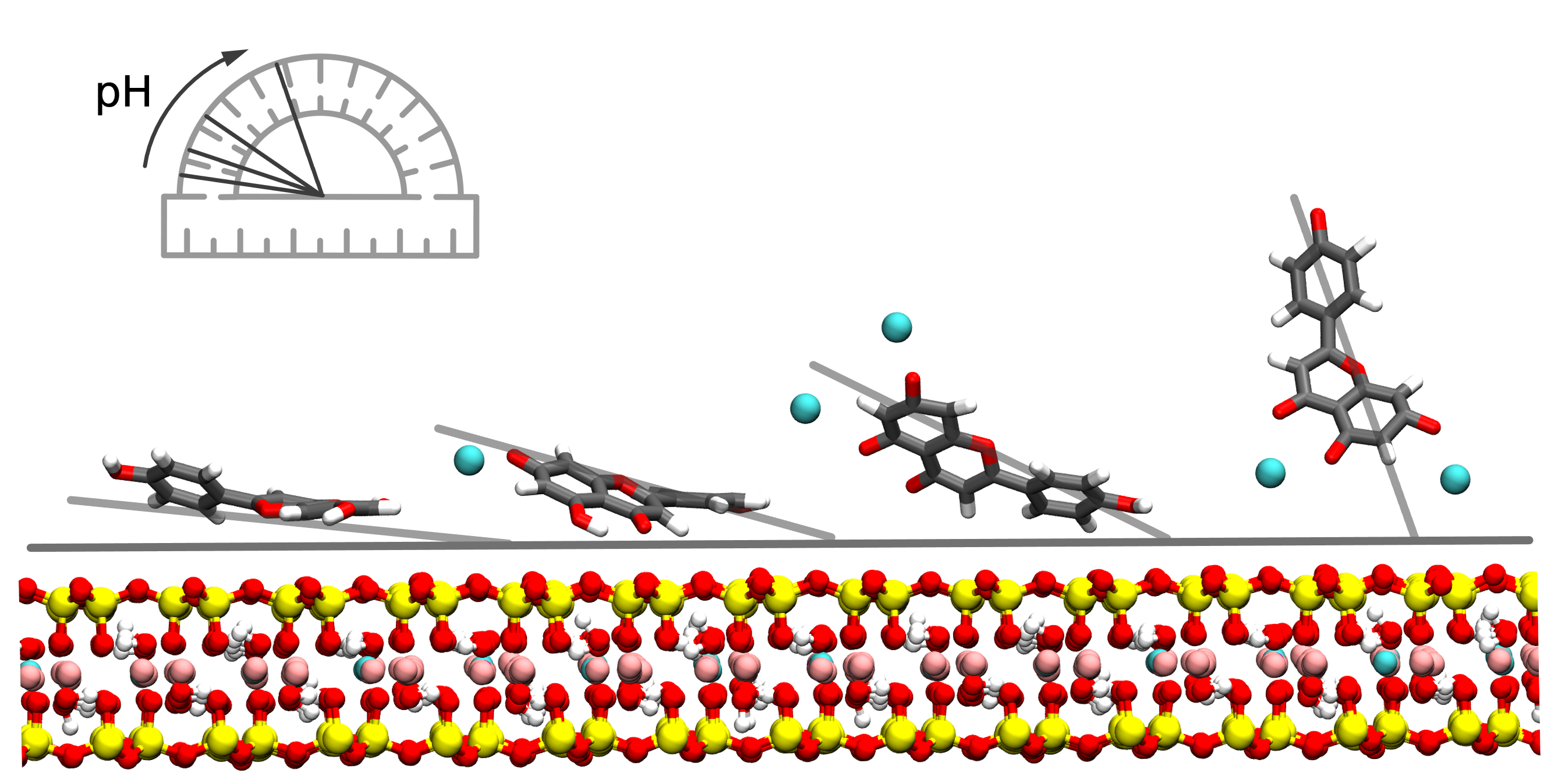}


\end{tocentry}

\begin{abstract}

Interactions between organic species and natural minerals are fundamental to the processes around us. 
With the aid of molecular dynamics simulations, we identify key adsorption mechanisms of apigenin on smectite clay minerals. 
The mechanism is highly sensitive to the pH -- changing from co-crystallisation in acidic-to-neutral solutions to the ion-bridging in mild-alkaline. 
The ionic species play a significant role in alkaline environments: the deprotonated apigenin species chelate metals, which, in turn, leads to the formation of a stable organic-metal-mineral complex and stronger adsorption in the presence of divalent cations.
Smectite clays buffer the solution to mildly alkaline; hence, the type of exchangeable cations in the clay will be critical in determining the adsorption mechanism and organic retention capacity. 
Overall, our study showcases a computational strategy that can be transferred to a wide variety of organic-mineral systems in the natural environment.

\end{abstract}


Interactions of organic molecules and minerals are the driving force behind many, seemingly unrelated, processes around us -- from soil signalling and nutrient retention, to applications for pollution control and remediation, to natural resource management, to pharmaceutical and biomedical applications, and even to the research into abiogenesis.\cite{uddin2017review, wang2022review, saha2020controlled, bosak2021searching, erastova2017mineral}
Nevertheless, identifying the mechanisms of these interactions, both in the laboratory and \emph{in silico}, is a difficult task owing to the shear variety of environmental factors and the complexity of the natural materials.\cite{liu2022molecular}

In this work, leveraging on the sampling through molecular dynamics simulations, we present a protocol for the study of interactions between apigenin, a flavonoid, and montmorillonite, a common clay mineral, under a range of environmental conditions. While representative models of organic molecules have become a norm in the modelling community, this is not the case for natural minerals. To this end, in our study, we ensure that the modelled montmorillonite clay is a truthful representation of its natural counterpart -- Wyoming montmorillonite, which is mined at one of the world's largest deposits at Newcastle formation, County of Crook, State of Wyoming, USA.\cite{sutherland2014wyoming} 
The molecular modelling study provides a clear understanding of how the interactions between organic molecules and natural clay minerals are affected by the changes in the surrounding environmental conditions. The molecular details obtained are informative to our understanding of a number of processes involving these species: from soil processes to the ageing of the natural pigments. 


\emph{Apigenin} is a polyphenolic secondary metabolite, abundant in flowers, herbs, fruits and vegetables. 
Along with other flavonoids, it has been used in traditional medicine, leading to a growing volume of research into their pharmaceutical properties, in particular, for the prevention of free-radical-mediated diseases, including cancer.\cite{medina2017apigenin, ginwala2019potential} 
In nature, flavonoids are poorly degraded in soils and are an important signalling molecule in the microbe-plant communication.\cite{mierziak2014flavonoids, cesco2012plant, del2020soil}  

Apigenin, like most flavonoids, is bright yellow colour and has been used as a natural wool and silk dye.\cite{deveouglu2019review, favaro2007acidichromism} Flavanoids are \emph{mordant-dyes}, i.e. require the fibres to be pre-treated with a \emph{mordant} -- a metal salt that binds to amino and carboxyl groups of the fibres and allows for the formation of a metal complex with the dye molecule, fixating it on the fibre.\cite{saunders1994light} 
Interestingly, flavonoids have also been suggested as the colouring agents in Maya Yellow, a dye-clay complex pigment.\cite{domenech2011maya, domenech2014isomerization}

\emph{Clay minerals} are layered silicates, often featuring isomorphous substitutions, that create a permanent negative layer charge, counterbalanced by hydrated interlayer cations. These cations form a flexible interlayer space that can swell to accommodate ions with larger hydration spheres and/or organic species. \emph{Smectites} are a group of swelling clays, commonly present in soils, and have been used anthropogenically for their known ability to \emph{smiktrís} (Greek for clean) contaminated materials.\cite{beneke2002fuller} 

Smectites' notable ability is the protection and preservation of intercalated organic species, at the same time offering their environmentally-controlled release. These organic-mineral interactions play a major role in the soils' nutrient delivery, soil-biosphere regulatory and signalling processes, as well as in the detection of molecular fossils in archaeological or extraterrestrial settings.\cite{bosak2021searching}

Yet, little is known about the fate of flavonoids in soils or their interactions with clay minerals, with only a limited number of studies, reporting heterogeneous and inconsistent observations.\cite{sosa2010persistence} To this end, with the aid of molecular dynamics simulations, we investigate the mechanism of adsorption of apigenin on montmorillonite under a range of environmental conditions, testing the effect of pHs and mono- and divalent cations \ce{NaCl} and \ce{CaCl2}, respectively.

\begin{figure}
\centering
\includegraphics[scale=0.6]{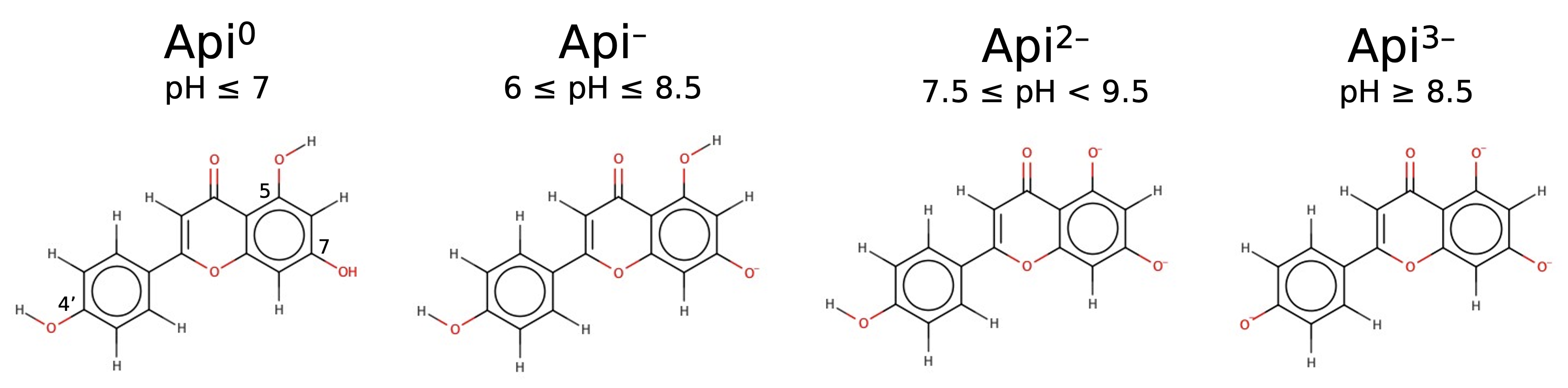}
\caption{Chemical structures of apigenin species at different protonation states, indicating the naming used in this work and the pH ranges where more than 15\% occurrence of the specie is observed. The atom numbering for hydroxyl groups is given over the neutral apigenin.}
\label{fig:api-protonation}
\end{figure}

 
Apigenin (Api), like other aglycone flavonoids, is poorly soluble in water, while its solubility increases significantly in mild alkaline environments.\cite{lucas2019effect, papay2016comparative} The acid-base dissociation constants for Api are close to each other, and the multi-species mixtures are often found, making the identification of pKa values difficult. Nevertheless, both experimental and computational studies agree that the hydroxyl group at C7 is most readily deprotonated (see Figure \ref{fig:api-protonation}), with the reported pKa$_1$ values in the range of 6.6 to 7.4.\cite {spiegel2022quantum, favaro2007acidichromism, medina2017apigenin} 
The next pKa$_2$ is between 8.1 and 9.3, and can be attributed to the deprotonation at C4’ or C5 positions, it is also possible that there is an overlap of two dianion forms.\cite {spiegel2022quantum, favaro2007acidichromism, medina2017apigenin} The last pKa$_3$, producing trianion, is rarely measured and is reported to be above 10.5, or even 11.5.\cite {spiegel2022quantum} In this work, we analyse interactions of four Api species -- a neutral form and three anions -- which are present at more than 15\% across the pH range. These species and the pH ranges are given in Figure \ref{fig:api-protonation}, while complete molar fraction distributions of Api species as a function of pH are given in the SI, Table S1.

The smectite clay chosen for this study is Wyoming Na-montmorillonite (SWy) -- a well-characterised and most commonly studied clay in the laboratory due to its easy purchase from the Clay Minerals Society. While the molecular simulations of montmorillonite clays have also been done frequently,\cite{yang2017atomistic, underwood2015eor, loganathan2019understanding, underwood2016ion} the models used feature idealised structure, not capturing the variety of isomorphic substitutions and of their positions, as found in the natural samples.
The natural SWy clay's structure is (Ca$_{0.12}$Na$_{0.32}$K$_{0.05}$) [Al$_{3.01}$Fe$_{0.41}$Mg$_{0.54}$Mn$_{0.01}$Ti$_{0.02}$] Si$_{7.98}$Al$_{0.02}$O$_{20}$(OH)$_4$.
We ensured that our molecular model of SWy MMT was as closely matching to the experimental SWy composition as mathematically possible for finite system size. We were able to set up the following structure: (Ca$_{0.11}$Na$_{0.343}$) [Al$_{3}$Fe$_{0.43}$Mg$_{0.57}$] Si$_{8}$O$_{20}$(OH)$_4$, ensuring that the isomorphic substitutions are random, but are not neighbouring. Since the valance of octahedral iron in clay may be both II and III, based on the known charge per unit cell, the SWy MMT only features \ce{Fe^{III}}. The SWy clay is a swelling clay with the reported interlayer spacing of 1.56 nm.\cite{holmboe2012porosity, kozaki2008diffusion}
Therefore, accounting for hydration shells of Na$^+$ and Ca$^{2+}$ cations, we used 12 water molecules per counter ion, which resulted in the desired interlayer spacing.\cite{yang2017atomistic, holmboe2012porosity}

\begin{figure}
\centering
\includegraphics[scale=0.45]{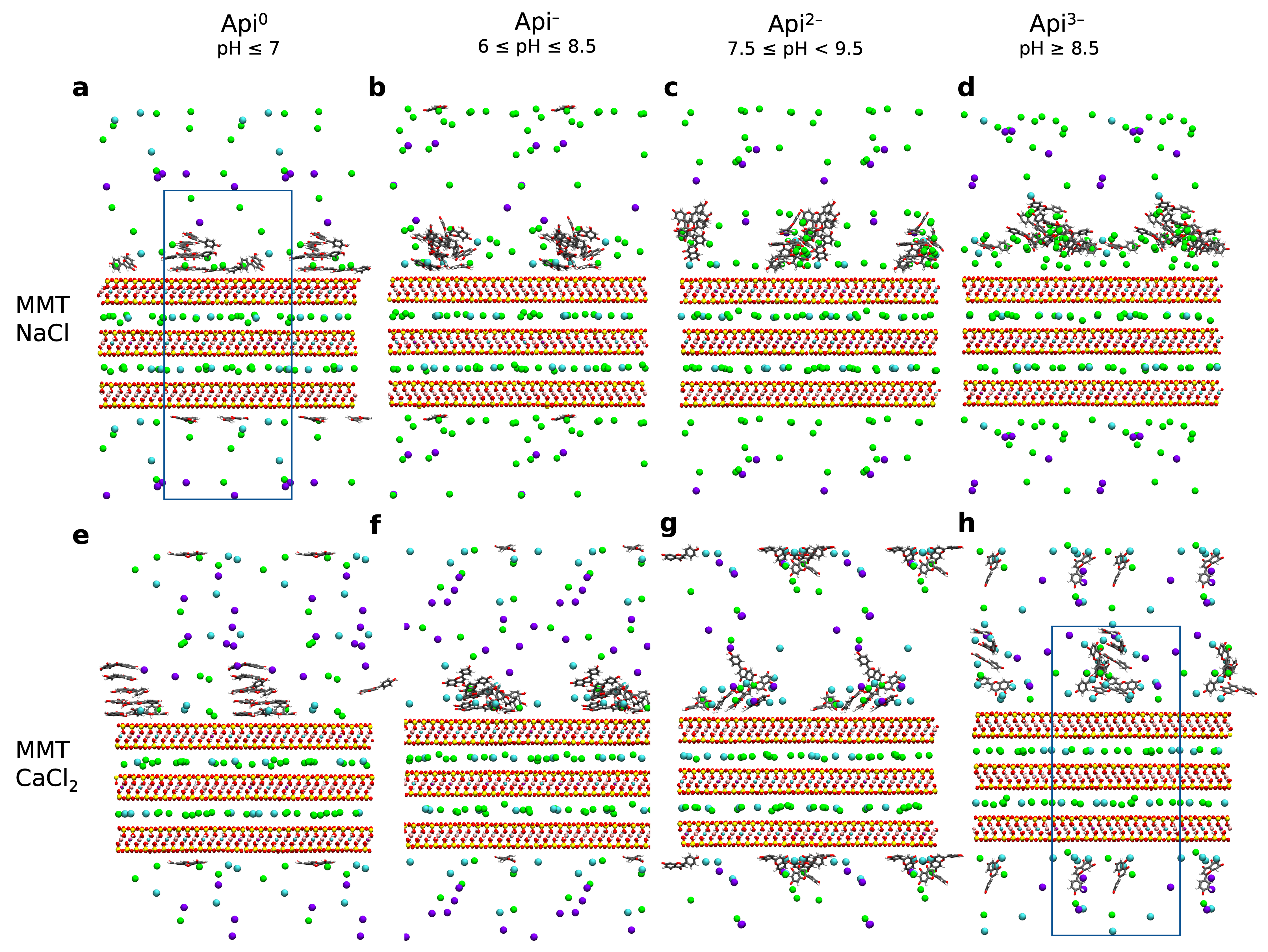}
\caption{Snapshots of apigenin-montmorillonite systems after 100 ns of simulation, varying the apigenin species according to the given pH ranges, and in the 0.1 M of \ce{NaCl} (top row) or \ce{CaCl2} (bottom row) salt solution. Water is not shown for clarity and a periodic simulation box is marked for the two example systems only.}
\label{fig:apimmt}
\end{figure}

We performed molecular dynamics simulations of the 0.2 M each of the four apigenin species in 0.1 M solutions of either \ce{NaCl} or \ce{CaCl2} and in the presence of MMT clay. The final structures are shown on Figure \ref{fig:apimmt}. Additionally, for reference, we carried out studies without MMT clay, and of MMT clay with the 0.1 M \ce{NaCl} or \ce{CaCl2} solutions only. The full details of the system set-up and simulation protocols are given in the Methods section, 
while the complete set of analyses and additional renderings of these simulations are given in the SI.

With nearly all molecules fully protonated under a pH of 6.5, Api$^0$ precipitates, forming orderly $\pi$-$\pi$ stacks with a crystalline structure. No effect of cations is observed. In the presence of clay, in order to minimise the surface area exposed to water, hydrophobic Api$^0$ precipitates onto the silicate surface with the aromatic rings parallel to the surface, as seen in Figure \ref{fig:apimmt}(a,e). The ionic composition of the surrounding solution does not affect this co-precipitation mechanism, as ions remain in the solution or continue to charge-balance the clay. These observations are quantified via the linear density profiles and radial distributions given in the SI. 

\begin{figure}
\centering
\includegraphics[scale=0.45]{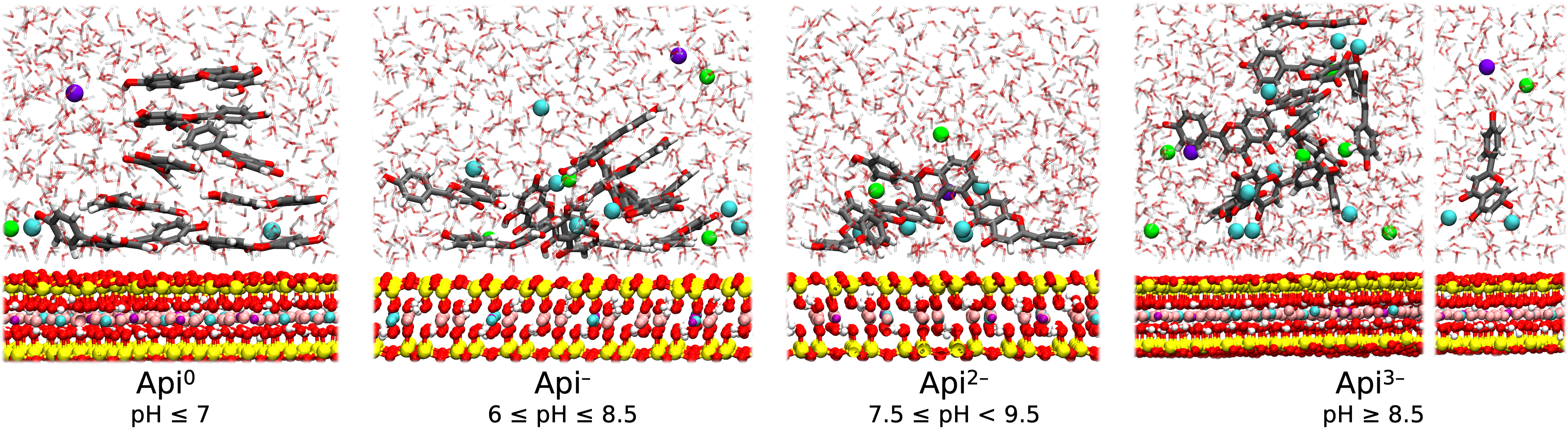}
\caption{Change in the adsorption mechanism of apigenin on the montmorillonite clay in the presence of CaCl$_2$ with change of pH.}
\label{fig:mechanism_apimmt}
\end{figure}

As the pH increases towards the first pKa, the deprotonation of C7-OH produces a single anionic form -- Api$^{1-}$. While in the experimental system, we should expect both non-dissociated and anionic apigenin species are found, with our model, we choose to examine the interactions of each anionic form independently. This Api$^{1-}$ remains poorly soluble, but unlike its neutral counterpart, the crystalline stacking is distorted by the cations coordinating to the negative hydroxy group.
This behaviour is also observed in the presence of MMT clay. The Api$^{1-}$ molecules are associating with the clay surface (Figure \ref{fig:apimmt}(b,f)), still forming stacks. Nevertheless, due to the repulsion of negative charges, the deprotonated C7-O$^{-}$ group points away, coordinating the cation. This rotates the stack to create a 20-degree tilt to the surface, see Figure \ref{fig:mechanism_apimmt}.

With further increase in the alkalinity of the system beyond pH = 8, apigenin deprotonates further to form dianion Api$^{2-}$. In solution, the molecules are more mobile; nevertheless, they still form $\pi$-$\pi$ stacked agglomerates, exposing the deprotonated groups towards the interface with water. 
When clay surface is available, apigenin molecules will interact with it, as shown in Figure \ref{fig:apimmt}(c,g). At this point, the valence of cations has a significant effect on the adsorption behaviour.
While at lower pHs, the cations were merely charge-balancing the system and hydrophobic interactions were guiding the adsorption; in the alkaline environment, cations form a Stern-like layer and shield the two interacting negative species -- apigenin and clay. The negative groups of Api$^{2-}$ are even stronger repelled from the clay, while the protonated hydroxy groups form H-bonds with the oxygens of the siloxane surfaces. This results in a 60-degree twist of the molecules to the surface. Apigenin stack is now running along the clay surface in a herringbone way, rather than away from the surface as seen for lower-charged species (Figure \ref{fig:mechanism_apimmt}). In the case of Ca$^{2+}$ cations, we notice apigenin is more mobile in the interlayer space, and the adsorption peak is higher nearer to the surface than in the presence of Na$^+$. This signifies a higher surface affinity in the presence of a higher valence counter ions.

With the slight further pH increase, fully deprotonated trianion Api$^{3-}$ can be found. In these forms, apigenin is increasingly more soluble in water, but still, the formation of agglomerates is observed.  
Interactions with clay can no longer rely on H-bonding, and therefore, the adsorbed Api$^{3-}$ is found further away from the surface than the first hydration shell. When the major counter ion is monovalent, the Na$^+$-Api$^{3-}$ agglomerate is poorly adsorbed and remains mobile (Figure \ref{fig:apimmt}(d,h)). In the presence of Ca$^{2+}$, the adsorption is via cationic bridging, with individual Api$^{3-}$ molecules interacting with two Ca$^{2+}$, coordinating to the deprotonated group at C5 and at C7 positions, while the C4’ hydroxyl group is pointing orthogonal to the surface (Figure \ref{fig:mechanism_apimmt}). Since our clay systems were set up to account for natural exchangeable counter ions, the small amount of Ca$^{2+}$, native to MMT, will facilitate apigenin adsorption even in the Na-dominated system. Furthermore, in the environment where Na$^+$ ions can be exchanged for divalent or even trivalent ions, the adsorption of apigenin will also be enhanced.

It has to be noted that clay minerals are also known for their buffering capability, further enhanced by the adsorption of metals. Typically, smectic soils will lead to slightly alkaline conditions, with pH ranges between 7.5 and 8.5.\cite{yong1990buffer, kittrick1971montmorillonite} 
Therefore, in the smectic soils or in the presence of smectite minerals, apigenin is expected to be in its deprotonated forms -- Api$^{-}$, Api$^{2-}$ and even some amount of Api$^{3-}$. In such cases, apigenin-clay interactions are strongly influenced by the presence and type of metal cations.\cite{favaro2007acidichromism, saunders1994light} 

Interestingly, the record of Ca$^{2+}$-flavone interactions can be traced back to the studies of mordants and natural pigments. The preparations needing to be alkaline, could combine the use of alum (acidic) and be balanced by alkaline ``chalk-containing clays''. It was noted that larger amounts of calcium would lead to less-stable dyes than those based on hydrated alumina only.\cite{saunders1994light} 

Furthermore, the described apigenin-metal complexation is attributed to the apigenin's antioxidant properties. The chelation of the redox-active metals, such as \ce{Cu^{II}} and \ce{Fe^{II/III}}, minimises the radical-generating Fenton reaction within the organism.\cite{spiegel2022quantum}

To this end, the described apigenin-clay interactions are in-line with our knowledge from experimental and colloquial observations. Further investigations of the chemistry of flavonoids and natural clay minerals, ensuring the truthful representation of their structure and accounting for naturally present metals and salts, will also inform us on the structure-property relations in Maya dyes, including Maya Yellow, legendary for their resistance to ageing, weathering and acid treatments. \cite {domenech2011maya, domenech2014isomerization} 

Furthermore, through this study, we demonstrated how molecular dynamics simulations give insights into the interactions of organic molecules and minerals, this approach is transferable to a wide range of natural systems.


\section{Computational Methods}

\subsection {Model Set-up}

Apigenin protonation states and molar fraction distribution have been calculated through MarvinSketch with Major Macrospecies and pKa plug-ins, Marvin version 22.20, ChemAxon. The molar fraction distributions for apigenin species are given in Table S1, SI. We have selected the species that are found in amounts of $>$15\% at any value of pH. Those species are: neutral Api$^0$, monoinionic Api$^{-}$, dianionic Api$^{2-}$ and trianionic Api$^{3-}$. Their structures are given in Figure \ref{fig:api-protonation}. The apigenin structures were assigned topology with CHARMM CGenFF force field \cite{vanommeslaeghe2012automation, vanommeslaeghe2010charmm} using an online interface at paramchem.org.

The clay mineral was developed to match the natural Wyoming Montmorillonite (SWy). SWy is a well-characterised clay, used in many studies in the literature. This natural clay has the following unit cell structure: (Ca$_{0.12}$Na$_{0.32}$K$_{0.05}$)  [Al$_{3.01}$Fe$_{0.41}$Mg$_{0.54}$ Mn$_{0.01}$Ti$_{0.02}$] Si$_{7.98}$Al$_{0.02}$O$_{20}$(OH)$_4$.
To create a model, representative of real SWy clay, we have made a number of unit cells that could be found within such clay, each having full atomic occupancy.  We then created a model system, combining these unit cells in a 7 $\times$ 5 $\times$ 3 in x, y, and z-directions. We ensured that in the final structure, the isomorphic substitutions were not neighbouring, the layer charges were in line with the natural mineral and that the hydration shell of the counter ions produced the experimental interlayer spacing of 1.56 nm.\cite{holmboe2012porosity, kozaki2008diffusion}
The model montmorillonite structure can be reduced to the following unit cell: (Ca$_{0.11}$Na$_{0.343}$) [Al$_{3}$Fe$_{0.43}$Mg$_{0.57}$] Si$_{8}$O$_{20}$(OH)$_4$. A detailed description of the MMT model is given in Table S2 and Figure S3 in the SI.
The system was assigned the ClayFF force field with the single-point charge, SPC, water.\cite{cygan2004molecular} We would like to note, that the octahedral iron in the original publication has been changed since the first publication, and so the parameter should be taken from the later review.\cite {cygan2021advances}

The ClayFF and CHARMM force fields are based on the same combination rules, making them compatible. The combined use of these force fields has been tested in our previous works.\cite{underwood2015eor, tian2018understanding, underwood2016wetting}

The MMT model set-up is a 2-dimensional periodic clay system with a 3.063077 $\times$ 4.50630 nm$^2$ surface. We have then expanded one of the interlayer spaces to 10 nm in the z-direction, creating the bulk space for adsorption studies. The space was populated with 10 apigenin species (resulting in ~0.2 M concentration), charge-balanced by either Na$^+$ or Ca$^{2+}$ ions, added background salinity of 5 \ce{NaCl} or \ce{CaCl2} (resulting in ~0.1 M concentration) and remaining space filled with water. It also has to be noted, that clay has exchangeable ions, which were present here as 12 Na$^+$ and 4 Ca$^{2+}$, according to the MMT structure. 
Furthermore, the control systems without MMT or without apigenin were also set up. All of the systems simulated in this work are summarised in Table \ref{tbl:Systems}. 

\begin{table}
  \caption{Composition of systems simulated}
  \label{tbl:Systems}
  \begin{tabular}{c|cccc}
    \hline
    System Name              & Clay      & No. of Api$^{q}$   & pH range                  & No. Ion$^{q}$, Salinity  \\
    \hline
    MMT Api$^0$ NaCl         & SWy MMT       &  10 Api$^0$       & pH$\leq$7.0             & 5 NaCl \\
    MMT Api$^{-}$ NaCl       & SWy MMT       &  10 Api$^-$       & 6.0$\leq$pH$\leq$8.5    & 10 Na$^+$, 5 NaCl \\
    MMT Api$^{2-}$ NaCl      & SWy MMT       &  10 Api$^{2-}$    & 7.5$\leq$pH$<$9.5       & 20 Na$^+$, 5 NaCl\\
    MMT Api$^{3-}$ NaCl      & SWy MMT       &  10 Api$^{3-}$    & pH$\geq$8.5             & 30 Na$^+$, 5 NaCl\\
    MMT Api$^0$ CaCl$_2$     & SWy MMT       &  10 Api$^0$       & pH$\leq$7.0             & 5 CaCl$_2$ \\
    MMT Api$^{-}$ CaCl$_2$   & SWy MMT       &  10 Api$^{1-}$    & 6.0$\leq$pH$\leq$8.5    & 5 Ca$^{2+}$, 5 CaCl$_2$ \\
    MMT Api$^{2-}$ CaCl$_2$  & SWy MMT       &  10 Api$^{2-}$    & 7.5$\leq$pH$<$9.5       & 10 Ca$^{2+}$, 5 CaCl$_2$\\
    MMT Api$^{3-}$ CaCl$_2$  & SWy MMT       &  10 Api$^{3-}$    & pH$\geq$8.5             & 15 Ca$^{2+}$, 5 CaCl$_2$\\
    \hline
    MMT NaCl                 & SWy MMT       &  --               & --                      &  5 NaCl \\
    MMT CaCl$_2$             & SWy MMT       &  --               & --                      &  5 CaCl$_2$ \\
    \hline
    Api$^0$  NaCl            & --            &  10 Api$^0$      & pH$\leq$7.0              & 5 NaCl \\
    Api$^{-}$ NaCl           & --            &  10 Api$^-$      & 6.0$\leq$pH$\leq$8.5     & 10 Na$^+$, 5 NaCl \\
    Api$^{2-}$ NaCl          & --            &  10 Api$^{2-}$   & 7.5$\leq$pH$<$9.5        & 20 Na$^+$, 5 NaCl\\
    Api$^{3-}$ NaCl          & --            &  10 Api$^{3-}$   & pH$\geq$8.5              & 30 Na$^+$, 5 NaCl\\
    Api$^0$  CaCl$_2$        & --            &  10 Api$^0$      & pH$\leq$7.0              & 5 CaCl$_2$ \\
    Api$^{-}$ CaCl$_2$       & --            &  10 Api$^-$      & 6.0$\leq$pH$\leq$8.5     & 5 Ca$^{2+}$, 5 CaCl$_2$ \\
    Api$^{2-}$ CaCl$_2$      & --            &  10 Api$^{2-}$   & 7.5$\leq$pH$<$9.5        & 10 Ca$^{2+}$, 5 CaCl$_2$\\
    Api$^{3-}$ CaCl$_2$      & --            &  10 Api$^{3-}$   & pH$\geq$8.5              & 15 Ca$^{2+}$, 5 CaCl$_2$\\
  \end{tabular}
\end{table}

\subsection {Simulation Protocol}
The simulations were performed with GROMACS 2022.3 molecular dynamics package.\cite{hess2008gromacs} Each simulation was first energy minimized using the steepest descent algorithm with the convergence criterion where the maximum force on any one atom is less than 500 kJ mol$^{-1}$ nm$^{-1}$. Then the systems were then equilibrated over 5 ns in the isothermal-isobaric (NPT) ensemble, using a velocity-rescale thermostat and a semi-isotropic Berendsen barostat set to 300 K and 1 bar, respectively. Using the Berendsen thermostat allowed for smooth equilibration of interlayer spacings. Subsequently, the production runs were carried out in the isothermal-isobaric ensemble with a velocity-rescale thermostat set at 300 K and semi-isotropic Parrinello-Rahman at 1 bar. The minimization and production simulations were run with real-space particle-mesh-Ewald (PME) electrostatics, and a Van der Waals cut-off of 1.4 nm. This follows our previously established protocols. \cite{underwood2015eor,tian2018understanding,underwood2016wetting, erastova2017mineral}

The production runs have been carried out for 100 ns for each of the eight MMT-Apigenin systems, for 50 ns for the eight apigenin-only reference systems and for 20 ns for MMT-only systems. These times were deemed sufficient after accessing the evolution of species through RMSD and DynDen\cite{degiacomi2021dynden}. 
 
\subsection {Analysis and Visualisation}
For the analysis, we have used the last half of the production runs, i.e. last 50 ns from the 100 ns simulations. last 25 ns for the 50 ns simulations and the last 10 ns for the 20 ns simulations. The analysis carried out were linear density profiles and radial distributions of species, those were done using the analysis tools within GROMACS 2022.3 package.\cite{hess2008gromacs} The data was plotted with Python 3.6 using the Matplotlib package.\cite{hunter2007matplotlib} All of the analysis data is given in the SI.

The snapshots were rendered using VMD 1.9.4,\cite{Humphrey1996vmd} with colours as follows: Apigenin atoms: C - grey, H - white, O - red; Ions: Na$^+$ - green, Ca$^{2+}$ - cyan, Cl$^-$ - purple; Clay: Si - yellow, O - red, H - white, Al - pink, Mg - cyan, Fe - green; water is always present, but may not be shown for clarity.

\begin{acknowledgement}

The authors thank French-Azerbaijani University's high-performance computing services and especially Gadir Rustamli for their assistance during the use of the facilities. Valentina Erastova would like to thank the Chancellor's Fellowship by the University of Edinburgh. 

\end{acknowledgement}

\begin{suppinfo}

Data tables and a complete analysis of molecular dynamics simulations, including additional renderings, are provided in the Supporting Information. 

\end{suppinfo}

\bibliography{bibliography}

\end{document}